\documentstyle[aps,prl,epsfig]{revtex}

\begin{document}

\draft \wideabs{ \title{The Stability Balloon for Two-dimensional Vortex
    Ripple Patterns}
  
  \author{J. L. Hansen $^{1,2}$, M. van Hecke$^1$, C. Ellegaard$^1$,
    K.  H. Andersen$^{1,3}$, T. Bohr$^4$, A. Haaning$^1$ and
    T. Sams$^2$}

  \address{$^1$ Niels Bohr Institute, Blegdamsvej 17, DK-2100 Copenhagen {\O}\\
    $^2$ Danish Defense Research Establishment, Ryvangs all\'e 1,
    Postbox 2715,
    DK-2100 Copenhagen {\O} \\
    $^3$ ISVA, Danish Technical University, building 115, DK-2800 Kgs.
    Lyngby\\ $^4$Institute of Physics, Danish Technical University, 
Building 309, DK-2800
    Kgs. Lyngby }
  
  \date{\today}
  
  \maketitle

\begin{abstract}
  Patterns of vortex ripples form when a sand bed is subjected to an
  oscillatory fluid flow. Here we describe experiments on the response
  of regular vortex ripple patterns to sudden changes of the driving
  amplitude $a$ or frequency $f$. A sufficient decrease of $f$ leads
  to a ``freezing'' of the pattern, while a sufficient increase of $f$
  leads to a {\em supercritical} secondary ``pearling'' instability.
  Sufficient changes in the amplitude $a$ lead to {\em subcritical}
  secondary ``doubling'' and ``bulging'' instabilities.  Our findings
  are summarized in a ``stability balloon'' for vortex ripple pattern
  formation.
\end{abstract}

}

\narrowtext

A flat bed of sand subjected to an oscillatory flow of water is seldom
stable but instead displays the formation of patterns.  Classical
studies \cite{bagnold} have shown that after the flat bed becomes
linearly unstable, so-called rolling grain ripples (small heaps of
grains) are formed first. These are, however, always transient
\cite{stegner}, and eventually strongly nonlinear {\em vortex ripples}
are formed in a coarsening type process. These ripples have triangular
crests with slopes roughly at the angle of repose, and the flow around
the ripple crests is dominated by vortices that occur in the wake of
the ripples.  Together with the converging flow at the ``wind'' side
of the ripples, these vortices yield sand mass transport directed
toward the crests of the ripples, which is balanced by sand
avalanching down when the slopes grow too large. The wavelength of
such ripple patterns is comparable to the amplitude of the fluid
motion which sets the scale for the size of the separation bubbles
\cite{stegner,nielsen,toymodel}; this wavelength is substantially larger than
the most unstable wavelength of a flat bed
\cite{blondeaux,andersen}.

To characterize the vortex ripples
we have studied their
{\em pattern forming} properties.  From this perspective, the system
combines a number of unique features. 
Firstly, the driving is anisotropic which results in alignment of the
ripples perpendicular to the flow. This allows for studies in
one-dimensional geometries \cite{stegner,scherer}, although our
experiments indicate that instabilities of the ripples lead to
intrinsically two dimensional patterns. 
Secondly, typical ripple
wavelengths are essentially independent of the system and grain
dimensions and $f$, but scale with the driving amplitude $a$
\cite{stegner,nielsen}.
Finally, due to the strongly nonlinear character of the development,
it has so far not been possible to describe the pattern dynamics
in terms of an ``amplitude equation''.

Our setup consists of a tray of sand that is oscillated with amplitude 
$a$ and frequency $f$
in a tank of water, allowing us to study two-dimensional patterns.
To probe the fully developed vortex ripples, we have studied their
response to sudden changes of the control parameters $a$
\begin{figure}
  \epsfxsize=.98\hsize \mbox{\hspace*{-.05\hsize}\vspace{-3mm}
    \epsffile{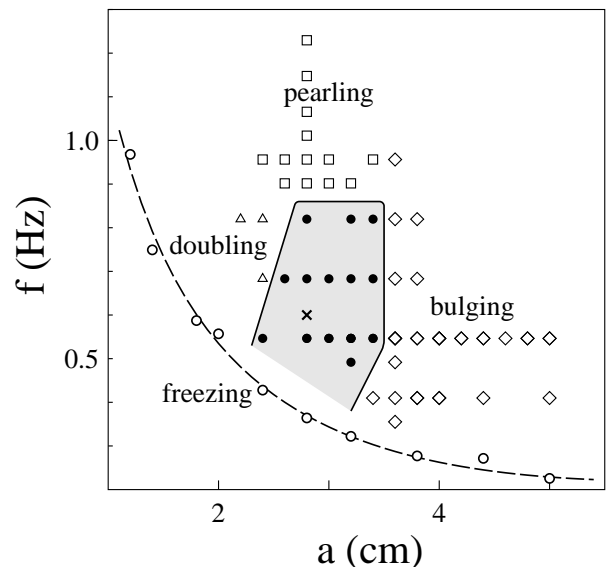}} \vspace{-1mm}
    \caption{Stability balloon for fully developed vortex ripples. The
    initial ripple pattern with wavelength $4.2$ cm is compatible with
    driving parameters $a\!=\!2.8$ cm and $f\!=\!0.6$ Hz (cross) and
    remains stable for drivings indicated in full circles. Triangles,
    squares and diamonds indicate drivings where this pattern
    experiences a doubling, pearling or bulging instability. The open
    circles correspond to the $n\!=\!10$ measurements shown in
    Fig.~\ref{grains-fig} and the dashed line roughly indicates where
    the pattern freezes.}  \label{fig0}
\end{figure}
\noindent  and $f$. 
Our findings lead to a ``stability balloon'' for
vortex ripple pattern formation shown in Fig.~\ref{fig0}.
First we study the amount of grain motion for
patterns that have evolved ``freely'' from the flat bed and have a
wavelength selected by the driving amplitude. 
This leads to the ``freezing line'' in Fig.~\ref{fig0} (and discussed in
more detail in Fig.~\ref{grains-fig}).
In the second set of experiments we study the response of regular
patterns with initially {\em fixed} wavelengths to changes in $f$ or
$a$.  We find that a secondary ``pearling'' instability (Fig.~\ref{pearl-fig})
occurs when
the frequency is sufficiently increased, while
secondary ``doubling''
\begin{figure}[t] 
  \epsfxsize=.95\hsize \mbox{\hspace*{-.04\hsize}\vspace{-3mm}
    \epsffile{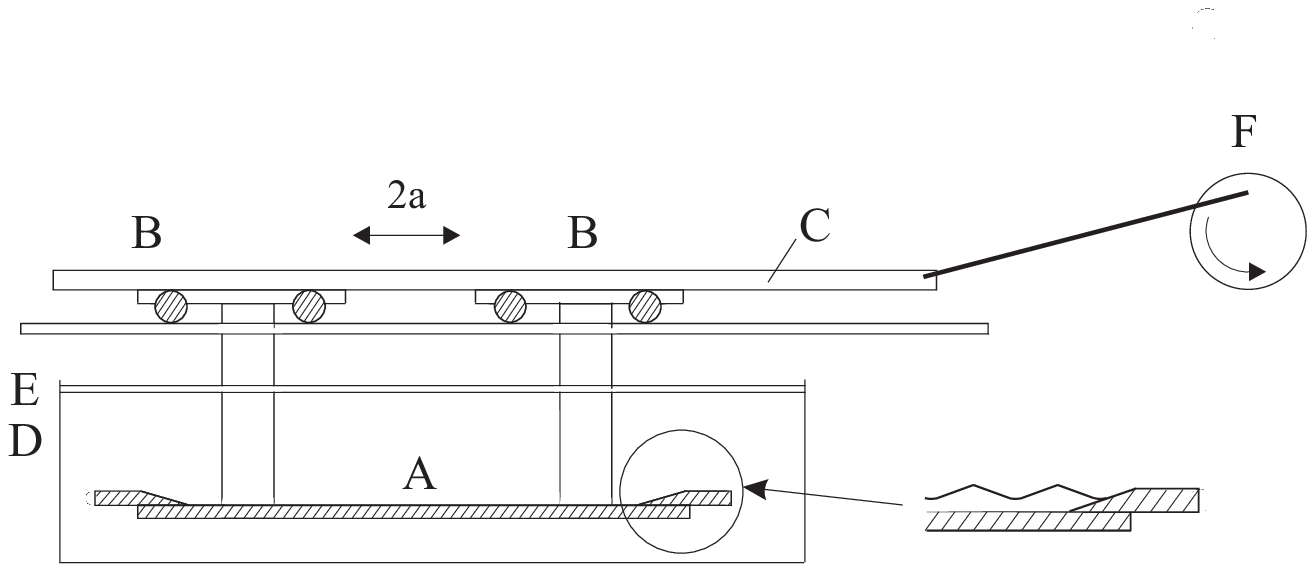}} \vspace{.2cm}
  \caption{Sketch of the experimental setup. The tray A is filled
with sand and its oscillations are driven by the motor F via the 
rail B.} \label{exp-fig}
\end{figure}

\noindent  and ``bulging'' (Fig.~\ref{pha-fig}) 
instabilities occur when the
amplitude $a$ is decreased or increased respectively, beyond some threshold.
A qualitative presentation of these instabilities has appeared in 
\cite{thwart}.  

{\em Experimental Setup -- } Our setup is sketched in
Fig.~\ref{exp-fig}.  An $0.6\,{\rm m}\times 1\,{\rm m}$ aluminum tray
(A) suspended from rollers (B) mounted on a stiff frame (C) is
immersed in a sufficiently deep tank (D) filled with water. To avoid
sloshing, a 3 cm thick flat plate of Plexiglas (E) is placed on top of
the tank.  The frame and tray are driven with a continuously
controllable frequency $f$ (period $T$) and amplitude $a$ by an ICME
AC motor (F). The sides of the tray consist of 2 cm high straight
boundaries, while the ends are triangular wedges with a slope of
$15^{\circ}$ and maximum height 2 cm (see Fig.~\ref{exp-fig}); the
rational behind these ``soft'' boundaries will be discussed below. The
``sand'' consists of spherical glass beads, ranging in sizes from 250
to 350 $\mu$m.  The thickness of the sand layer is smallest in the
troughs of the ripple pattern, but always larger than 5 mm.  The whole
setup is illuminated from the left side and filmed from above by a
Dalsa 8 bit CAD4 CCD camera with $1024^2$ resolution. A trigger is
mounted on the motor so that all pictures are taken at the same
extremal position of the tray. For the values of the driving
considered here (Fig.~\ref{fig0}), typical ripple patterns consists of
15-20 ripple lengths. Suspension can be ignored and the maximal
acceleration of the tray is well below the fluidization threshold. For
appropriate values of $a$ and $f$, fairly regular ripple patterns grow
from the flat bed. For example, for $a\!=\!2.8$ cm we find ripple
patterns with wavelength 4.2 cm.

{\it Grain motion --} Once a fully developed pattern is formed, how
does the number of grains in motion vary with $f$ and $a$?  We
observed that, due to irregularities in some grains, sand bed images
usually display a number of very bright spots. The difference between
two subsequent images is dominated by the appearance or disappearance
of a number $n$ of such bright spots; we assume that $n$ is
proportional to the number of grains which have moved.

To measure $n$ we proceed as follows: We start from a flat bed and
obtain an equilibrated pattern by running the system for one hour, so
that the driving amplitude selects the ripple wavelength, in contrast
to the secondary instability experiments presented below.  We then
take
\begin{figure}[t]
  \epsfxsize=1.2\hsize \mbox{\hspace*{-.08\hsize} \epsffile{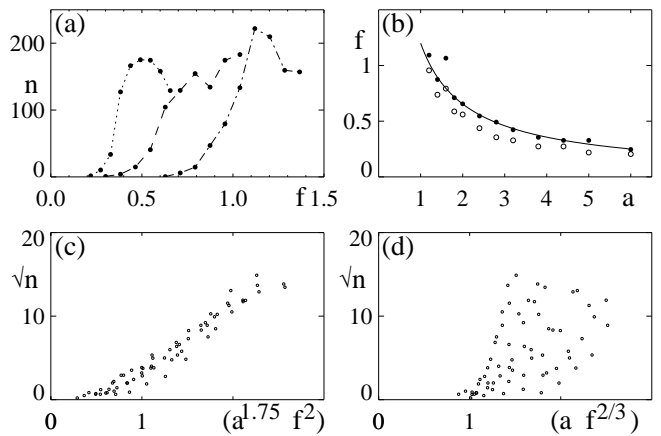}}
  \vspace{-1.8cm} \caption{(a) The average amount of moving grains $n$
  as a function of the driving frequency $f$ for amplitudes 1.4
  (dot-dashed), 2.4 (dashed) and 4.4 (dots). (b) Points in control
  parameter space where $n$ is 10 (open circles) and 50 (closed
  circles) compared to a curve where $a^{7/4} f^2$ is constant. 
(c-d) Data collapse of
  $\sqrt{n}$ versus the turbulent expression $ a^{7/4} f^2$ and the
  laminar expression $ A f^{2/3}$, with $a$  in cm and $f$ in Hz (see
  \protect\cite{shieldsnote}). }\label{grains-fig}
\end{figure}
\noindent a series 50 images from which the average of $n$ is
determined.  The frequency is then lowered to a new value, the system
is allowed to relax for fifteen minutes and fifty new images are
recorded. This procedure is repeated for ten decreasing values of the
frequency.  In Fig.~\ref{grains-fig}a, three examples of the number of
moving grains as function of frequency are shown. For high frequencies
$n$ saturates, and the corresponding data points were discarded in the
analysis presented below; this leaves us with a total of 65 data
points for 10 different values of $a$.

 We wish to obtain a relation between $a$ and $f$ where $n$ is
constant.  It was previously suggested that the maximum
non-dimensional shear stress on a flat bed, $\sigma_m$ would
be a good measure of the forcing
\cite{nielsen,toymodel,shieldsnote}. If the boundary layer were
laminar one would obtain that $\sigma_m \propto a f^{2/3}$ from the
solution to Stokes second problem \cite{landaulifshitz}, while for
turbulent flow a semi-empirical relation $\sigma_m \propto a^{1.75} f^2$
would be appropriate \cite{find_original_paper_ken}.  In
Fig.~\ref{grains-fig}c we have plotted $\sqrt{n}$ versus the turbulent
expression $a^{1.75} f^2$ and obtain a fairly good linear relation,
while the correlation between $n$ and the laminar expression $a
f^{2/3}$ is very weak (Fig.~\ref{grains-fig}d).

{\em Secondary instabilities -- } So far we have described patterns
with a wavelength that is selected by the driving amplitude. Now we
ask what happens when a perfectly regular pattern is suddenly
subjected to changes in the driving parameters $a$ or $f$.

Obtaining a completely regular pattern is not entirely trivial.
Freely grown patterns contain defects, which may be annihilated only
after
long times (on the order of days), although other distortions of the
pattern may then appear. These are partially driven by a small drift
of the pattern, which has a velocity on the order of one 
\begin{figure}[b]
  \epsfxsize=1.45\hsize \mbox{\hspace*{-.13\hsize} \epsffile{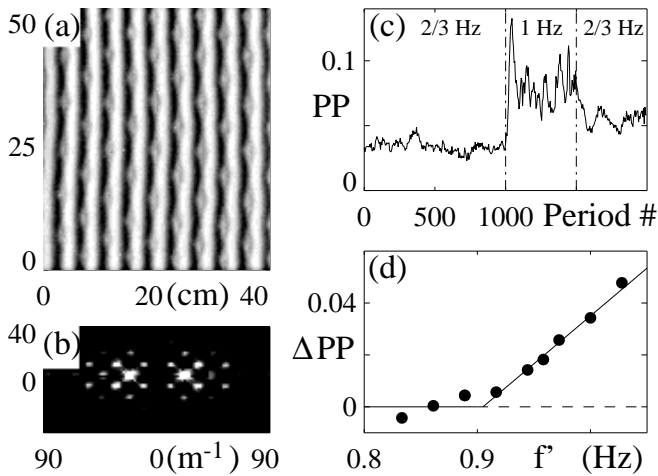}}
  \vspace{-2.8cm} \caption{(a) Central section (43 x 53 cm) of a
  pearling state obtained by subjecting a pattern with wavelength 4.2
  cm to a driving of $a\!=\! 2.8$ cm and $f\!=\!1$ Hz. (b)
  Corresponding power spectrum (range 188 x 94 cm$^{-1}$) (c)
  Time series of $PP$ when $f$ is changed from $2/3$ Hz to $f'\!=\!1$Hz
  and back again.  (d) The order parameter $\Delta PP$ as a function
  of the frequency $f'$.  For details see text.}  \label{pearl-fig}
\end{figure}
\noindent ripple wavelength per $10^4$ oscillations
\cite{drift-note}. The deformations due to the drift are partly
eliminated by using the slanted boundaries, allowing ripples to
``drift out'' of the system (leading to a small loss of sand at the
edge of the plate). To obtain completely regular initial conditions,
we have adopted the following procedure: First a flat bed is obtained
by fluidizing the sand during a short period of strong oscillations of
the plate.  Then a regular pattern is imprinted into the sand by
pressing down a frame with parallel equally spaced metal ridges.
Small irregularities are then eliminated by a few ($\approx10$)
oscillations of the plate.

By making large changes in $a$ or $f$ it is relatively simple to get
regular secondary instability patterns as shown in
Fig.~\ref{pearl-fig} and \ref{pha-fig}.  A study of the precise nature
of these transitions involves runs performed at parameters close to
the instability boundaries, and here the time scale for the
development of instabilities diverges.  As a result, the ``ideal''
dynamics may get hidden behind experimental artifacts.  By introducing
appropriate order parameters we will below precisely characterize both
the pearling and the bulging instability.

{\em Pearling -- } When the driving frequency is increased beyond a
certain critical value, we find a secondary ``pearling'' instability
(Fig.~\ref{pearl-fig}). Here, the crests of the initial ripples remain
essentially undisturbed, but in their troughs new small ripples
(pearls) emerge, and the resulting pearling pattern is periodic with
the pearls aligned on lines inclined by approximately 45 degrees.  
The pearling transition is a
{\em supercritical} secondary instability.
When pearls are formed, their
strength quickly saturates at some well defined value, and they
disappear when the frequency is 
lowered as shown in Fig.~\ref{pearl-fig}c:
Starting from 
\begin{figure}
  \epsfxsize=1.13\hsize \mbox{\hspace*{-.06\hsize} \epsffile{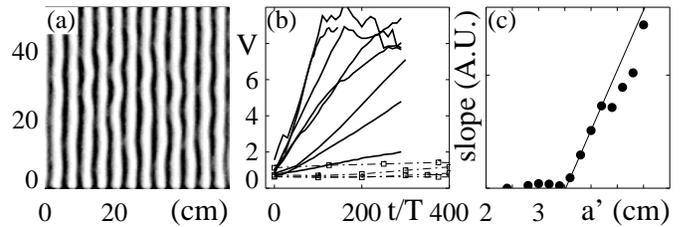}}
  \vspace{-4.1cm} \caption{(a) Central part (53 x 53 cm) of a bulging
  pattern, 525 T after the driving amplitude is changed to 4 cm
  ($f=0.41$).  (b) The growth of deviations of a straight ripple
  pattern, quantified by the variance $V$ for $f\!=\!0.55~$Hz and a
  range of amplitudes. The thin curves with symbols correspond to
  amplitudes $2.4,2.8,3.0,3.2$ and $3.4$ cm (below the bulging
  instability), while the thick curves correspond to
  $3.6,3.8,4.0,4.2,4.4,4.6,4.8$ and $5.0~$cm (above the bulging
  instability).  (c) The maximum slope of $V$ (for times up to $300
  T$) as a function of $a$ shows a sharp transition for $a\approx
  3.55$ cm.}  \label{pha-fig}
\end{figure}
\noindent an imprinted perfect ripple pattern with 
wavelength $4.2$ cm (corresponding to an amplitude
$a\!=\!2.8$ cm), the system
was driven at a low frequency of 0.67 Hz.  After 1000 oscillations,
the frequency was suddenly increased to $f'$, kept
there for 500 oscillations and then again lowered to the initial
frequency. 

To characterize the strength of the pearls, we have measured $PP$, the
total intensity in the primary satellite peaks of the power spectrum
(Fig.~\ref{pearl-fig}b). Due to finite size effects and noise, $PP$ is
not zero for perfect patterns, and as order parameter we therefore use
$\Delta PP$, the difference between mean values of $PP$ during period
1000-1500 and period 0-1000.  A plot of $\Delta PP$ as a function of
the quenching frequency $f'$ shows a a well-defined transition point,
above which the order parameter increases continuously (see
Fig.~\ref{pearl-fig}d).  A further increase of the frequency leads to
more erratic states and finally to fluidization where the ripple
patterns are washed away.

{\em Bulging: -- } When the driving amplitude
$a$ is increased sufficiently, the regular ripple patterns become
unstable to two dimensional modulations.  While we cannot completely
rule out that this is a long wavelength instability
\cite{CrossHohenberg93}, our data strongly suggests that the
wavelength of this modulation in the longitudinal direction is locked
on four times the wavelength of the underlying pattern, with a similar
wavelength in the transversal direction. In contrast to the pearling
instability, this instability is subcritical.  The bulging
deformations grow until neighboring ripples become so close that they
form defects, which climb and glide rapidly through the system,
finally leading to a regular pattern with a larger wavelength.
For sequence of pictures showing the development of the
instability see \cite{thwart}.

Close to the instability boundary it becomes very difficult to
distinguish slow drift from slow development of the instability, and
we have therefore developed the following sensitive measure for the
onset of the instability.  Starting with a perfect pattern of
wavelength 4.2 cm, we suddenly shift the amplitude.  To characterize
the time evolution of the pattern, we have extracted the local values
of the ``longitudinal'' ripple length $\lambda_i$ taken over the whole
two-dimensional image. The variance $V$, defined as $\Sigma_{i=1}^N
(\lambda_i- \bar{\lambda})^2 /(N-1)$, is then a simple measure for the
amount of deformations in the pattern. The evolution of $V$ is shown
in Fig.~\ref{pha-fig}b for a variety of values of the amplitude $a$.

Even below the formation of bulges, $V$ grows slowly due to slow large
scale deformations of the pattern, although the growth rate is
essentially independent of the driving amplitude $a$.  For $a$ above
some critical value, $V$ displays a clear, surprisingly linear,
increase with time. In Fig.~\ref{pha-fig}c we have plotted the maximum
of $(V(t f +100)-V(t f))/100$ for $t$ up to period 200.  This
quantity, which measures the maximum slope of $V$, clearly identifies
the location of the secondary instability at $a\!=\!3.55(5)$ cm.  In
this way we can distinguish between slow expansion or contraction of
the pattern and the bulging instability.

{\em Doubling -- } When the amplitude $a$ 
is decreased sufficiently, a sub-critical
{\em doubling} instability occurs. 
The initial phase of the development of this instability suggests that
it can be captured in a one-dimensional framework. When the driving
amplitude gets sufficiently small, the separation vortices that drive
the sand transport do no longer reach over the through between
ripples.  This leads to the formation of bumps in the ripple troughs,
which in turn grow out to form new ripples. Behavior similar to this
has been seen in numerical studies of 1D vortex ripple patterns
\cite{toymodel,kenphd}.  For a picture of this transition see
\cite{thwart}.

{\em Discussion and outlook -- } We have characterized some of the
pattern forming properties of fully developed vortex ripple
patterns. By observing the number of moving grains on the ripples we
have showed that the maximum shields parameter on the flat bed,
calculated using the turbulent expression, is a relevant control
parameter. We have shown that regular vortex ripple patterns are
stable to small changes in the driving amplitude and frequency. When
these changes become too large, however, the vortex ripples show a rich
variety of secondary instabilities; pearling, bulging and doubling.

Vortex ripples pose many theoretical challenges. The existence of a
stable band and the doubling transition have been found in a simple
model of ripple patterns \cite{toymodel}.  The bulging and pearling
are, to the best of our knowledge, not present in any simple
theoretical models, such as Swift-Hohenberg type models incorporating
local mass conservation, left right symmetry and a finite wavelength
instability (which leads to spatial derivatives of order six). We
believe that the origin of the instabilities is basically
hydrodynamical and related to the dynamics of the separation zones. In
the doubling transition it is thus clearly seen in the 1-d experiments
that the new ripples originate approximately at the reconnection point
for the separation vortex. The bulging and pearling transitions are
genuinely 2-d and thus more complicated, but we speculate that the
bulging transition is basically a Rayleigh-Plateau ``sausage"
instability of the almost cylindrical separation vortex, whereas the
pearling instability might be related to the centrifugal instability
of the cylinder, giving rise to transverse Taylor vortices.  Obviously
these ideas need considerable elaboration in view of the strong time
dependent shear experienced by the separation vortex.

It is a pleasure to acknowledge discussions with M.-L. Chabanol,
J. Krug, A.  Stegner and E. Wesfreid.


\begin{thebibliography}{10}
\bibitem{shieldsnote} The Shields parameter on a flat bed is defined
as $\tau/( g d (\rho_s-\rho_w))$, where $\tau$ is the maximal shear
stress on the bed, $g$ denotes gravity, $d$ is the diameter of the
sand grains and $\rho_s$ and $\rho_w$ are the densities of the sand
and water. For laminar flow the maximal shear stress is given by $U
\sqrt{2\pi f \nu}$ where $U$ is the maximal fluid velocity, $f$ the
frequency and $\nu$ the viscosity. For turbulent conditions, $\tau$
has been fitted to a numerical solution as $0.02 \rho_w f U^2
(a/k_N)^{-0.25}$, where $a$ is the fluid amplitude and $k_N$ is the
Nikuradze roughness of the bed \cite{find_original_paper_ken}.  The
number of moving grains is related to the Shields parameter as $n =
6/(\pi d^2 \mu) (\sigma - \sigma_c)$, where $\sigma_c$ is the
threshold of motion and $\mu$ the dynamical friction coefficient.
\bibitem{drift-note} When one attempts to stop the drift by inserting
  steep ridges at each end of the moving sand tray, slow compression
  and dilatation of the patterns near these boundaries occurs. For
  such boundaries we frequently observe formation of defects after
  long time, even when $a$ is chosen such as to accommodate the
  imprinted pattern best.
\end{thebibliography}
\end{document}